\documentclass{iopart}
\usepackage{pstricks,graphicx,amsbsy,bbm,amsmath,graphicx}
\begin{document}
\title{Input-output relations for multiport ring cavities}
\author{Matteo G. A. Paris}
\address{Dipartimento di Fisica dell'Universit\`a
di Milano, Italia}
\begin{abstract} 
Quantum input-output relations for a generic $n$-port ring 
cavity are obtained by modeling the ring as a cascade of $n$
interlinked beam splitters.  Cavity response to a beam impinging
on one port is studied as a function of the beam-splitter
reflectivities and the internal phase-shifts. Interferometric
sensitivity and stability are analyzed as a function of the number 
of ports.
\end{abstract}
Multiport ring cavities represent the natural generalization of
two-port Fabry-Perot interferometers \cite{old} to several modes of the
radiation field. They find application in advanced
interferometry, division  multiplexing and optical cross connect.
Recently, a three-port fiber ring laser was suggested and
demonstrated to improve sensing resolution \cite{png}, whereas
multiport optical circulators have been used for interconnecting
single-fiber bidirectional ring networks \cite{jeo}. In addition,
a three-port reflection grating was demonstrated \cite{sch} and
the corresponding input-output relations have been derived
\cite{thr}.  From a more fundamental perspective, multiport
couplers, either multiport beam splitters or ring cavities, are
crucial devices to generate and engineer multiphoton entangled
states \cite{alm}. In fact, the cavity response is linear in 
the input modes for both kind of devices, with ring cavities 
offering the additional feature of a high nonlinearity with 
respect to the internal phase-shifts of the cavity.
The use of ring cavities, supplemented by nonlinear media, 
has been also suggested to realize nondemolitive measurement
and photon filtering \cite{fkf}.
\par
In this letter, fully quantum input-output relations for a generic 
$n$-port ring cavity are obtained by modeling the ring as a cascade 
of $n$ interlinked, suitably matched, beam splitters.  
In this way, the cavity response to an impinging beam, as 
well as the use of the cavity in interferometry, can be evaluated 
as a function of the beam-splitter reflectivities, the 
internal phase-shifts, and the number of ports.
\par
\begin{figure}[h!]
\centerline{\includegraphics[width=0.5\textwidth]{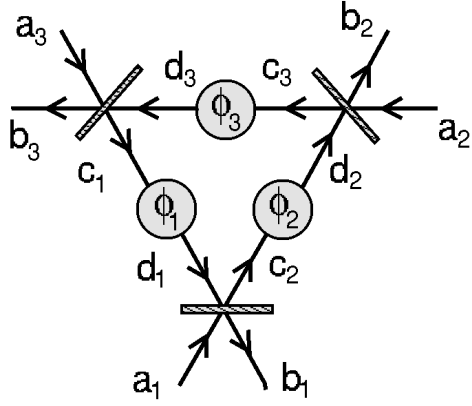}}
\caption{\label{f:three} Three-port ring cavity as a cascade 
of three interlinked beam splitters. The cavity is built by 
three beam splitters with equal transmissivity $\tau$. The 
matching relations $d_k= e^{i\phi_k}\:c_k$  should be 
satisfied together with the input-output 
relations (\ref{bs}) for each beam splitter.}
\end{figure}

Let us first illustrate the results in details for the case 
of a three-port ring cavity, which has been schematically 
depicted in Fig.  \ref{f:three}.
We assume that the three beam splitters used 
to build the cavity have the same transmissivity $\tau$. We also 
assume that losses at the beam splitters are negligible. The 
reflectivity of each coupler is thus given by $\rho=1-\tau$.
The input-output relation for the three beam splitters
are given by 
\begin{align}\label{bs}
BS_k: \quad \left\{\begin{array}{cl}
b_k &= \tau^{\frac12} d_k + \rho^{\frac12} a_k
\\ c_{1\oplus k} &= - \rho^{\frac12} d_k + \tau^{\frac12} a_k
\end{array}\right.
\end{align}
with $k=1,2,3$ and $\oplus$ denoting sum modulo $3$.
Any additional phase-shift at the beam splitters 
may be absorbed into the internal phase-shifts $\phi_k$.
In order to build the cavity the matching relations 
$d_k= e^{i\phi_k}\:c_k$  should be also satisfied, 
together with  Eqs. (\ref{bs}). After lengthy but 
straightforward calculations one arrives at the 
input-output relations for the cavity 
\begin{align}\label{trs}
b_1 &= \frac{1}{A_3}\left\{\sqrt{\rho}\left[1+\sqrt{\rho}e^{i\phi}\right]
a_1 - \tau \sqrt{\rho} e^{i\phi_{13}} a_2 + \tau e^{i\phi_1} a_3\right\}
\nonumber \\ 
b_2 &= \frac{1}{A_3}\left\{\tau e^{i\phi_2} a_1+ \sqrt{\rho}\left[1+\sqrt{\rho} 
e^{i\phi}\right] a_2 - \tau \sqrt{\rho} e^{i\phi_{12}} a_3 \right\}
\nonumber \\ 
b_3 &= \frac{1}{A_3}\left\{- \tau \sqrt{\rho} e^{i\phi_{23}} a_1 + \tau 
e^{i\phi_3} a_2 + \sqrt{\rho}\left[1+\sqrt{\rho}e^{i\phi}\right] a_3 \right\}
\:,
\end{align}
where $\phi_{jk}=\phi_j+\phi_k$, $\phi=\phi_1+\phi_2+\phi_3$ and 
$ A_3 = 1 + \rho^\frac32 e^{i\phi}$.
Unitarity of the mode transformations (\ref{trs}) can be explicitly 
checked through the normalization of the output modes $[b_j,b^\dag_k]=\delta_{jk}$.
\par\noindent
\par
\begin{figure}[h]
\centerline{\includegraphics[width=0.4\textwidth]{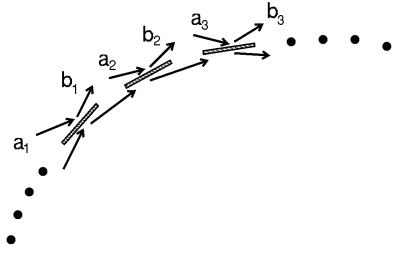}}
\caption{\label{f:N} Multiport ring cavity as a cascade of 
interlinked beam splitters.}
\end{figure}

The above model can be generalized to a cavity with an arbitrary number 
of ports, see Fig. \ref{f:N}. We have 
\begin{align}\label{modeN}
b_k^{(n)}= \sum_{j=1}^n M_{kj}^{(n)} a_j
\end{align}
where
\begin{align}\label{Matrix}
M_{11}^{(n)}=&\frac{1}{A_n}
\sqrt{\rho}\left[1+\: (-)^{1+n} \rho^{\frac{n}{2}-1} e^{i\phi}\right]
\nonumber \\
M_{1j}^{(n)}=&\frac{1}{A_n}\left\{
(-)^n \: \tau \: 
\rho^{\frac{n}{2}} \sum_{j=2}^n (-)^j \rho^{-\frac{j}{2}} 
e ^{i\theta_{1j}^{(n)}} \right\}
\end{align}
and cyclic transformations for $M_{kj}$, $k=2,...,n$, 
with $\phi=\sum_{k=1}^n \phi_k$, 
$A_n = 1 + \rho^{\frac{n}2} (-1)^{1+n} e^{i \phi}$ 
and $\theta_{tj}^{(n)}=\phi_t + \sum_{k=j+1}^n \phi_k$.
\par
Explicitly, for a four-port cavity we have
\begin{align}
b_1^{(4)}=\frac{1}{A_4}&\Big\{
\sqrt{\rho}\left[1-\rho e^{i\phi}\right]a_1 + \tau \rho e^{i\phi_{134}}
a_2 - \nonumber \\ &   
\tau \sqrt{\rho} e^{\phi_{14}} a_3 + \tau e^{i\phi_1} a_4 \Big\}
\end{align}
and cyclic transformations, where $\phi_{134}\equiv
\theta_{12}^{(4)}=\phi_1+\phi_3+\phi_4$.
\par
\begin{figure}[h]
\centerline{\includegraphics[width=0.5\textwidth]{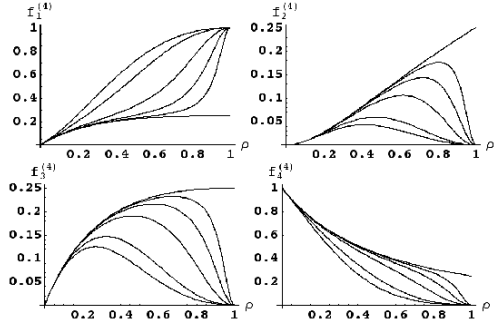}}
\caption{\label{f:f4} 
Cavity responses $f_k^{(4)}$, $k=1,..,4$ of a four-port ring-cavity 
as a function of the mirrors' reflectivity for different values 
of the total internal phase-shift. In plot of $f_1^{(4)}$ 
($f_k^{(4)}$ for $k\neq 1$), 
from bottom to top (from top to bottom) the curves corresponding 
to $\phi=0,\frac{\pi}{20}, \frac{\pi}{10}, \frac{\pi}{5}, 
\frac{\pi}{2},\pi$ respectively. }
\end{figure}

Let us now consider the situation in which one of the port (say, 
port 1) is fed by a coherent beam $|\alpha\rangle$, whereas the 
other ports are left unexcited.
Using mode transformations (\ref{modeN}) one may analyze the cavity 
response to a given excitation, {\em i.e.} how the input mean energy 
$\langle a^\dag_1 a_1 \rangle \equiv \langle \alpha| a^\dag_1 a_1|\alpha 
\rangle = |\alpha|^2$ is distributed
among the $n$ output photocurrents $I_k^{(n)}=b^{(n)\dag}_k b_k^{(n)}$, 
$k=1,...,n$ obtained by detecting light at the $n$ output ports 
of the cavity.
Being the mode transformations linear also the output 
beams are coherent state with amplitudes 
$|\beta_k^{(n)}\rangle$.
Upon defining the cavity response as 
$$f_k^{(n)}(\rho,\phi) = \frac{\langle b_k^{\dag(n)} b_k^{(n)}\rangle}{
\langle a_1^{\dag(n)} a_1^{(n)}\rangle}\:, $$
one has $\beta_k^{(n)}= \alpha\sqrt{f_k^{(n)}} \exp\{i \theta_k^{(n)}\}$
with $\theta_k^{(n)}=\arg M_{k1}^{(n)}$ and 
\begin{align}
f_1^{(n)} & =   \frac{\rho}{|A_n|^2} \left[ 1+ \rho^{n-2} + 2\:
(-)^{1+n} \cos\phi \: \rho^{\frac{n}{2}-1} \right] \\ 
f_k^{(n)} & = \frac{(1-\rho)^2}{|A_n|^2}\: \rho^{n-k} 
\qquad 2 \leq k \leq n
\end{align}
where $|A_n|^2= 1 + \rho^n + 2\: (-)^{1+n} \rho^{\frac{n}{2}}\cos\phi\: $.
The cavity response explicitly depends on the 
mirror reflectivity $\rho$, while, remarkably, it depends on the internal 
phase-shifts $\phi_k$ only through the total phase-shift $\phi$. 
The following sum-rule holds 
$$f_1^{(n)} + \sum_{k=2}^n f_k^{(n)}=1  \qquad
\forall n,\:\forall \phi,\:\forall\rho \:, $$ 
which, in turn, assures energy conservation.
In Fig. \ref{f:f4} we show the cavity responses 
$f_k^{(n)}$ for $n=4$ as a function of the mirror
reflectivity for different values of 
the total internal phase-shift. Notice that 
$0\leq f_k^{(4)}\leq 1$
for $k=1,4$ and 
$0\leq f_k^{(4)}\leq \frac14$ otherwise.
\par
In general, the minimum  of the cavity response at the first port 
is achieved for $\phi=0$ for $n$ even and for $\phi=\pi$ for $n$ odd.
For these values (cavity at resonance) we have 
\begin{align}
f_1^{(n)} &= \rho \left(\frac{1-\rho^{\frac{n}{2}-1}}{1-
\rho^{\frac{n}{2}}}\right)^2
\\
f_k^{(n)} &= \rho^{n-k} \left(\frac{1-\rho}{1-\rho^{\frac{n}{2}}}\right)^2
\quad 2 \leq k \leq n
\label{minfs}\;,
\end{align}
either for $n$ even or odd.
In the high-reflectivity limit $\rho\rightarrow 1$ we 
have $f_1^{(n)}=(1-\frac{2}{n})^2$ and 
$f_k^{(n)}=\frac{4}{n^2}$ $\forall k\neq 1$. 
In  other words, in a two-port cavity at resonance the energy is completely 
transferred to the second mode, while increasing the number of ports the 
energy is unavoidably ``more distributed''. For 
large $n$ the input beam is mostly reflected on
the beam $b_1^{(n)}$ and the cavity becomes opaque. 
An equal distribution at the output is obtained 
for $n=4$ ($f_k^{(4)}=\frac14$).
The cavity response $f_1^{(n)}$ at the first port (last port $f_n^{(n)}$
respectively) monotonically increases (decreases) as the mirror reflectivity
approaches unit value. On the other hand, $f_k^{(n)}$, $\forall k\neq 1,n$ 
show a maximum value, whose location depends on the internal phase-shift, as 
well as the number of ports of the cavity.
\par
\begin{figure}[h]
\centerline{\includegraphics[width=0.5\textwidth]{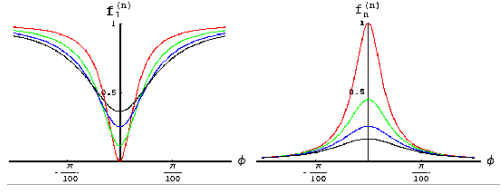}}
\caption{\label{f:finesse} 
Cavity responses $f_1^{(n)}$ (left) and 
$f_n^{(n)}$ (right) for a 
multiport ring cavity as a function of the internal 
phase-shift $\phi$. The responses for $n=2,3,4,5$ and $\rho=0.99$ 
are reported. The smaller is $n$, the peaked are the curves.}
\end{figure}

The sensitivity of the cavity in detecting perturbations
to the internal phase-shift decreases as the  number of ports 
increases. This is true either monitoring the cavity output
at resonance or doing the same at a fixed working point in an 
interferometric setup. 
In Fig. \ref{f:finesse} we show 
the cavity responses$f_1^{(n)}$ and $f_n^{(n)}$  
as a function of the internal phase-shift for different
number of ports. As it is apparent from the plot 
the curves flatten as the number of ports increases.
The full-width half-minimum (maximum) of $f_1^{(n)}$  ($f_n^{(n)}$), 
for a generic value of $n$, is given by
\begin{eqnarray}
\delta\phi_{\hbox{\tiny HW}}^{(n)}\simeq \frac{1-\rho^{n/2}}{2\rho^{n/4}} 
\stackrel{\rho\rightarrow 1}{\simeq} \frac{n}{4} (1-\rho)
\label{fwhm}\;,
\end{eqnarray}
showing a linear increases of the half-width.
\par
\begin{figure}[h]
\centerline{\includegraphics[width=0.35\textwidth]{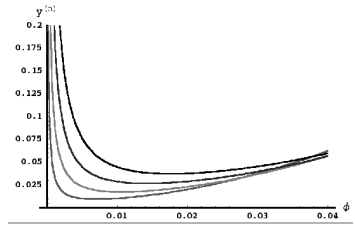}}
\caption{\label{f:int} Rescaled sensitivity $y^{(n)} = |\alpha| 
\delta\phi^{(n)}$ as function of working point 
$\phi$ for $\rho=0.99$ and 
for different values of $n$. 
From bottom to top $n=2,3,4,5$.}
\end{figure}

More generally, if one aims to detect the fluctuations
of the internal phase-shift around a fixed working point 
$\phi=\phi^*$ by monitoring the output photocurrents 
$I_k^{(n)}$ then the minimum detectable fluctuation corresponds
to the quantity \cite{shb}
\begin{eqnarray}
\delta\phi_k^{(n)} = 
\left|\left(\frac{\delta 
\langle I_k^{(n)}\rangle}{\delta\phi}\right)_{\phi=\phi^*}\right|^{-1}
\sqrt{\langle \Delta I_k^{(n)2}\rangle} 
\label{df}\;, 
\end{eqnarray}
where $\langle\Delta I_k^{(n)2}\rangle=\langle 
(b_k^{(n)\dag }b_k^{(n)})^2 
\rangle - \langle b_k^{(n)\dag }b_k^{(n)}\rangle^2$ 
denote the rms fluctuations of the output photocurrents.
When a single input port is excited in a coherent state 
$\alpha\rangle$ also the output signals are coherent and 
Eq. (\ref{df}) rewrites as 
\begin{eqnarray}
\delta\phi_k^{(n)} = \frac{\sqrt{f_k^{(n)}} 
}{|\alpha|}\: 
\left|\left(\frac{\partial f_k^{(n)}}{\partial\phi}
\right)_{\phi=\phi^*}\right|^{-1}
\label{df1}\;,
\end{eqnarray}
where $|\alpha|$ corresponds to the square root of the 
incoming average number of photons.
The optimal working point $\phi^*$,
corresponding to maximum sensitivity, is the internal phase-shift
that minimizes the value of
$\delta\phi_k^{(n)}$ 
We found that $\phi^*$ is close, but not equal, to 
$\phi=0$ for $n$ even and to $\phi=\pi$ for $n$ odd.
Only slight differences are observed for different values
of $k$, which vanishes for in the high-reflectivity regime. 
As a matter of fact, by increasing $n$ the optimal
working point $\phi^*$ moves away from $\phi=0$ ($\phi=\pi$) and
the minimum value of $\delta\phi$ increases. 
Since in the high-reflectivity regime $\rho \rightarrow 1$ 
the quantities $\delta\phi_k^{(n)}$ do not depends on $k$ at 
fixed $n$, {\em i.e} $\delta\phi_k^{(n)} = \delta\phi_1^{(n)}$, 
$\forall k=1,...,n$ the overall sensitivity of the cavity 
may be evaluated as $\delta\phi^{(n)}
=\delta\phi_1^{(n)}/\sqrt{n}$.
In Fig. \ref{f:int} we report the rescaled sensitivity 
$y^{(n)} = |\alpha| \delta\phi^{(n)}$ as function of $\phi$ 
for $\rho=0.99$ and for different values of $n$. As it is 
apparent from the plot the overall sensitivity slightly 
degrades with increasing $n$, despite the factor $1/\sqrt{n}$
decreases.
The curves versus $\phi$ flatten for increasing $n$ and 
this implies that the need of tuning of the cavity at the 
optimal working point also becomes less stringent, {\em i.e}
stability slightly increases.
\par
In conclusion, by modeling a $n$-port ring-cavity as 
a cascade of $n$ interlinked beam splitters we obtained 
its input-output relations in terms of the involved modes of 
the quantized radiation field. 
Using this approach, the cavity response to an impinging beam, 
as well as sensitivity to perturbations, can be straightforwardly 
evaluated as a function of the beam splitters reflectivity and 
the internal phase-shifts.
We found that increasing the number of ports the input energy
is unavoidably distributed over the output ports. The sensitivity 
of the cavity in detecting fluctuations of the internal phase-shift, 
either at resonance or at a fixed optimal working point, slightly 
degrades as the number of ports 
increases while, on the contrary, stability slightly increases.
\\ $ $ \\
This work has been supported by MIUR through the project 
PRIN-2005024254-002. The author thanks Maria Bondani for
several discussions.

\end{document}